\documentstyle[11pt,amssymb,cite]{article}

\def\ben{\begin{equation}}
\def\een{\end{equation}}

   \let\i=\iota 
  \let\n=\nu

\let\C=\Chi

\def\nn{\nonumber} \def\bd{\begin{document}} \def\ed{\end{document}}
\def\ds{\documentstyle} \let\fr=\frac \let\bl=\bigl \let\br=\bigr
\let\Br=\Bigr \let\Bl=\Bigl
\let\bm=\bibitem
\let\na=\nabla
\let\pa=\partial \let\ov=\overline
\newcommand{\be}{\begin{equation}}
\newcommand{\ee}{\end{equation}}
\def\ba{\begin{array}}
\def\ea{\end{array}}
\def\ft#1#2{{\textstyle{{\scriptstyle #1}\over {\scriptstyle #2}}}}
\def\fft#1#2{{#1 \over #2}}
\def\del{\partial}
\def\vp{\varphi}
\def\sst#1{{\scriptscriptstyle #1}}
\def\oneone{\rlap 1\mkern4mu{\rm l}}
\def\td{\tilde}
\def\wtd{\widetilde}
\def\ie{\rm i.e.\ }
\def\dalemb#1#2{{\vbox{\hrule height .#2pt
        \hbox{\vrule width.#2pt height#1pt \kern#1pt
                \vrule width.#2pt}
        \hrule height.#2pt}}}
\def\square{\mathord{\dalemb{6.8}{7}\hbox{\hskip1pt}}}
\newcommand{\ho}[1]{$\, ^{#1}$}
\newcommand{\hoch}[1]{$\, ^{#1}$}
\newcommand{\bea}{\begin{eqnarray}}
\newcommand{\eea}{\end{eqnarray}}
\newcommand{\ra}{\rightarrow}
\newcommand{\lra}{\longrightarrow}
\newcommand{\Lra}{\Leftrightarrow}
\newcommand{\ap}{\alpha^\prime}
\newcommand{\bp}{\tilde \beta^\prime}
\newcommand{\tr}{{\rm tr} }
\newcommand{\Tr}{{\rm Tr} }
\def\0{{\sst{(0)}}}
\def\1{{\sst{(1)}}}
\def\2{{\sst{(2)}}}
\def\3{{\sst{(3)}}}
\def\4{{\sst{(4)}}}
\def\5{{\sst{(5)}}}
\def\6{{\sst{(6)}}}
\def\7{{\sst{(7)}}}
\def\8{{\sst{(8)}}}
\def\n{{\sst{(n)}}}
\def\cA{{{\cal A}}}
\def\cB{{{\cal B}}}
\def\cF{{{\cal F}}}
\def\cH{{{\cal H}}}
\def\tV{\widetilde V}
\def\tW{\widetilde W}
\def\tH{\widetilde H}
\def\tE{\widetilde E}
\def\tF{\widetilde F}
\def\tA{\widetilde A}
\def\im{{i}}
\def\tY{{{\wtd Y}}}
\def\ep{{\epsilon}}
\def\vep{{\varepsilon}}
\def\R{\rlap{\rm I}\mkern3mu{\rm R}}
\def\bD{{{\bar D}}}

\def\R{\rlap{\rm I}\mkern3mu{\rm R}}
\def\bD{{{\bar D}}}
\def\R{{{\Bbb R}}}
\def\C{{{\Bbb C}}}
\def\H{{{\Bbb H}}}
\def\CP{{{\Bbb C}{\Bbb P}}}
\def\RP{{{\Bbb R}{\Bbb P}}}
\def\Z{{{\Bbb Z}}}
\def\bA{{{\Bbb A}}}
\def\bB{{{\Bbb B}}}
\def\bC{{{\Bbb C}}}
\def\bD{{{\Bbb D}}}
\def\bE{{{\Bbb E}}}
\def\bZ{{{\Bbb Z}}}
\def\Re{{{\frak{Re}}}}
\def\Im{{{\frak{Im}}}}
\def\cosec{{\,\hbox{cosec}\,}}
\def\Gm{{\Gamma_{\!\! -}}}
\def\Gp{{\Gamma_{\!\! +}}}
\def\stan{{standard }}
\def\nonstan{{supernumerary }}
\def\FF2{{ {}_{\sst 2}F_{\sst 1} }}

\newcommand{\tamphys}{\it Center for Theoretical Physics,
Texas A\&M University, College Station, TX 77843}

\newcommand{\upenn}{\it Department of Physics and Astronomy,\\ University
of Pennsylvania, Philadelphia, PA 19104}

\newcommand{\brussels}{\it Physique Th\'eorique et Math\'ematique,
Universit\'e Libre de Bruxelles,\\ Campus Plaine C.P. 231, B-1050
Bruxelles, Belgium}

\newcommand{\damtp}{\it DAMTP, Centre for Mathematical Sciences,
 Cambridge University,\\  Wilberforce Road, Cambridge CB3 OWA, UK}

\newcommand{\auth}{Z.-W. Chong{\hoch{\ddagger}}, G.W. Gibbons{\hoch *}, 
H. L\"u{\hoch{\ddagger}} and C.N. Pope{\hoch{\ddagger}} }

\begin{document}
\begin{flushright}

MIFP-04-09\ \ \ DAMTP-2004-56\\
{\bf hep-th/0405061}\\
May\  2004
\end{flushright}

\vspace{10pt}

\begin{center}

{\large {\bf Separability and Killing Tensors in Kerr-Taub-NUT-de
Sitter Metrics in Higher Dimensions}}

\vspace{20pt}
\auth

\vspace{20pt}

{\hoch{\ddagger}}{\it George P. \& Cynthia W. Mitchell
Institute for Fundamental Physics,\\ Texas A\& M University,
College Station, TX 77843-4242, USA}

\vspace{10pt}

{\hoch *}\damtp


\vspace{40pt}

\underline{ABSTRACT}
\end{center}

   A generalisation of the four-dimensional Kerr-de Sitter metrics to
include a NUT charge is well known, and is included within a class of
metrics obtained by Plebanski.  In this paper, we study a related
class of Kerr-Taub-NUT-de Sitter metrics in arbitrary dimensions $D\ge
6$, which contain three non-trivial continuous parameters, namely the
mass, the NUT charge, and a (single) angular momentum.  We demonstrate the
separability of the Hamilton-Jacobi and wave equations, we construct a
closely-related rank-2 St\"ackel-Killing tensor, and we show how the
metrics can be written in a double Kerr-Schild form.  Our results 
encompass the case of the Kerr-de Sitter metrics in arbitrary
dimension, with all but one rotation parameter vanishing.  Finally, we
consider the real Euclidean-signature continuations of the metrics,
and show how in a limit they give rise to certain recently-obtained
complete non-singular compact Einstein manifolds.

{\vfill\leftline{}\vfill \vskip 10pt \footnoterule 
{\footnotesize
Research supported in part by DOE grant
DE-FG03-95ER40917.}

}

\pagebreak
\setcounter{page}{1}

\newpage

\section{Introduction}

   Four-dimensional solutions of the Einstein equations have been
extensively studied for many decades.  In relatively recent times,
since the discovery of supergravity and superstring theory, solutions
of the Einstein equations, or the coupled Einstein-matter equations,
in higher dimensions have also been found to be of physical interest.
An encyclopaedic classification of all the known four-dimensional
solutions can be found in \cite{stkrmahohe}, but the higher-dimensional
cases have been less extensively investigated.  Various classes of
higher-dimensional solution have been obtained, including black holes
that generalise the four-dimensional Schwarzschild,
Reissner-Nordstr\"om and Kerr solutions.

   There are few general methods available for solving the Einstein
equations.  Almost always, one is forced to make a symmetry
assumption.  If the isometry group has orbits of codimension one, the
problem then reduces to solving non-linear ordinary differential
equations.  A much more challenging task results if the orbits have
higher codimension, since then non-linear partial differential
equations are involved.  Two techniques have proved useful in the past
for tackling cases like these, where the number of independent
variables is two or greater.  One technique is to adopt the
Kerr-Schild ansatz, which in effect reduces the Einstein equations to 
linear equations, and this has proved useful recently in obtaining
the general higher-dimensional version of the Kerr-de Sitter metrics
\cite{glpp}.  Another technique, pioneered by Carter, is to require of
the metric that it admit separation of variables for the
Hamilton-Jacobi equation, or for the wave equation or Laplace equation
\cite{carter1}.  This has the further advantage that having obtained
the metric, one is actually in a position to do something with it;
namely, to study its geodesics and eigenfunctions explicitly.

   In a remarkable paper, Carter exploited this idea to obtain a
general class of metrics in four dimensions which include the
Kerr-Taub-NUT-de Sitter solutions \cite{carter1}.  This general class,
in the formalism given by Plebanski \cite{pleb}, takes the form
\be
ds^2 = \fft{p^2 + q^2}{X}\, dp^2 +
\fft{p^2 + q^2}{Y}\, dq^2 +
\fft{X}{p^2 + q^2}\, (d\tau + q^2\,d\sigma)^2 -
\fft{Y}{p^2 + q^2}\, (d\tau -p^2\, d\sigma)^2\,,
\label{plebmetric0}
\ee
where
\be
X=\gamma -g^2 -\epsilon\, p^2 - \lambda\, p^4 +
2\ell\, p \,,\qquad
Y=\gamma +e^2 +\epsilon\, q^2 - \lambda\, q^4 -
2 m\, q\,.
\ee
They are solutions of the coupled Einstein-Maxwell equations with a
cosmological constant $\lambda$, and with electric and magnetic
charges given by $e$ and $g$.  We shall restrict attention to the case
of pure Einstein metrics in this paper, and so we set $e=g=0$.  The
remaining constants $(\gamma, m,\ell, \ep)$ effectively comprise 3
real continuous parameters and one discrete parameter, since one can
always make coordinate scaling transformations to absorb the magnitude
of, say, the dimensionless constant $\ep$.  Thus one may view
$(\gamma,m,\ell)$ as continuous parameters, and take $\ep=+1$, $-1$ or
0.  The constants $(\gamma,m,\ell)$ are related to the angular
momentum, mass and NUT charge.  Special cases of (\ref{plebmetric0})
include the Kerr-de Sitter solution and the Taub-NUT-de Sitter
solution.

   These four-dimensional metrics have a simple higher-dimensional
generalisation \cite{klemm}, which, as we shall show, also has the
property that both the Hamilton-Jacobi equation and the wave equation
may be solved by separation of variables.  Thus, for these very
special metrics, which encompass the Kerr-de Sitter metrics in
arbitrary dimension with all but one rotation parameter vanishing, the
geodesic flow on the cotangent bundle is a completely integrable
system in the sense of Liouville.  Such integrable dynamical systems
are comparitively rare, and when arising from the metric, are
associated with the existence of special tensor fields, called
St\"ackel-Killing tensor fields, on the manifold.  We shall construct
these explicitly for the higher-dimensional metrics.  Although the
S\"ackel-Killing tensor in the four-dimensional metrics admits a
Yano-Killing tensor square root, it appears rather unlikely that this
feature will extend to the higher-dimensional generalisations.

   A feature of the four-dimensional metrics is that they 
may be, upon analytic continuation to $(2,2)$ metric signature, be
cast into a double Kerr-Schild form, and this has played some r\^ole
in their construction, and that of more generalised metrics, by
Plebanski.  We find that the higher-dimensional metrics may also be
cast in double Kerr-Schild form, and that, although it is not true in
general, that the double Kerr-Schild form renders the Einstein equations
linear, in our case we find that it does result in linear equations.

   An important application of the original four-dimensional metrics
was the first construction of a complete, non-singular, compact,
inhomogeneous Einstein manifold \cite{page}.  This was done by analytically
continuing the metrics to positive-definite (Euclidean) signature, and
making a careful study of regularity conditions near coordinate
singularities.  We show that the same can be done for the
higher-dimensional metrics discussed in this paper.

\section{Higher-dimensional Generalisation}

   In this paper, we consider higher-dimensional generalisations of
the form
\be
d\hat s^2 = \fft{p^2 + q^2}{X}\, dp^2 +
\fft{p^2 + q^2}{Y}\, dq^2 +
\fft{X}{p^2 + q^2}\, (d\tau + q^2\,d\sigma)^2 -
\fft{Y}{p^2 + q^2}\, (d\tau -p^2\, d\sigma)^2 +
\fft{p^2\,q^2}{\gamma}\, d\Omega_k^2\,,
\label{plebmetric}
\ee
where
\be
X=\gamma -\epsilon\, p^2 - \lambda\, p^4 +
2\ell\, p^{1-k}\,,\qquad
Y=\gamma +\epsilon\, q^2 - \lambda\, q^4 -
2 m\, q^{1-k}\,,\label{xy}
\ee
and $d\Omega_k^2=g_{ij}\, dx^i\, dx^j$ is an Einstein metric on a
space of dimension $k$, normalised so that its Ricci tensor satisfies
$R_{ij} = (k-1)\, g_{ij}$.  One might, for example, take $d\Omega_k^2$
to be the metric on the unit sphere $S^k$.  It was shown in
\cite{klemm} that these metrics satsify the $D=k+4$ dimensional 
Einstein equation
\be
\hat R_{MN} = (k+3)\,\lambda\, \hat g_{MN}\,.\label{einst}
\ee
 
   The verification of the Einstein equations can be performed rather
straightforwardly in a coordinate basis.  From (\ref{plebmetric}), and
decomposing the coordinate indices as $M=(\mu,i)$, we may write the
components of the $D$-dimensional metric as
\be
\hat g_{\mu\nu} = g_{\mu\nu}\,,\qquad \hat g_{ij} = \fft{p^2\,
  q^2}{\gamma}\, g_{ij}\,,\qquad \hat g_{\mu i}=0\,,
\ee
where $g_{\mu\nu}$ is the four-dimensional Plebanski-type metric with
the modified functions $X$ and $Y$ given in (\ref{xy}).  It is easily
seen that the non-vanishing components of the affine connection $\hat
\Gamma^M{}_{NP}$ are given by
\bea
&&\hat\Gamma^\mu{}_{\nu\rho} = \Gamma^\mu{}_{\nu\rho}\,,\qquad
\hat\Gamma^i{}_{jk} = \Gamma^i{}_{jk}\,,\nn\\
&&\hat\Gamma^i{}_{j\mu} = \delta^i_j\, \del_\mu \log(p\, q) = 
\delta^i_j\, (\fft1{q}\, \delta^1_\mu + \fft1{p}\, \delta^2_\mu)\,,\\
&& \hat\Gamma^\mu{}_{ij}  = -\fft{p\, q}{\gamma\, (p^2+q^2)}\, 
(p\, Y\, \delta^\mu_1 + q\, X\, \delta^\mu_2)\, g_{ij}\,.\nn
\eea
Here, the explicit index values 1 and 2 refer to the $q$ and $p$
coordinates respectively.  From these expressions, it is
straightforward to substitute into the expression for the curvature,
and hence to verify that (\ref{plebmetric}) satisfies (\ref{einst}).

   The arbitrary-dimensional Kerr-de Sitter metrics with a single
rotation parameter $a$, which were obtained in \cite{hawhuntay}, arise
as special cases of the more general Einstein metrics
(\ref{plebmetric}).  Specifically, if we take the parameters in
(\ref{xy}) to be
\be
\gamma=a^2\,,\qquad \ep= 1- \lambda\, a^2 \,,\qquad m=M\,,\qquad \ell=0\,,
\ee
and define new coordinates according to
\be
p=a\, \cos\theta\,,\qquad q=r\,,\qquad \tau = t -\fft{a}{\Xi}\,
\phi\,,\qquad \sigma = - \fft{1}{a\, \Xi}\, \phi\,,
\ee
where $\Xi\equiv 1+\lambda\, a^2$, then (\ref{plebmetric}) reduces 
precisely to the metrics obtained in \cite{hawhuntay}.  The more
general solutions that we have obtained include the NUT charge
$\ell$ as an additional non-trivial parameter, when $D\ne 5$.

   The case $D=5$ is somewhat degenerate in the above construction,
in that the ostensibly additional NUT parameter $\ell$ is
fictitious in this case.  This is easily seen from the expressions
for the metric functions $X$ and $Y$ in (\ref{xy}) when $k=1$:
\be
X=\gamma -\epsilon\, p^2 - \lambda\, p^4 +
2\ell\,,\qquad
Y=\gamma +\epsilon\, q^2 - \lambda\, q^4 -
2 m\,,\label{xyd5}
\ee
One can absorb the parameter $\ell$ by means of additive shifts in
the constants $\gamma$ and $m$.  Since a constant 
scaling of the $S^k$ metric $d\Omega_k^2$ in (\ref{plebmetric}) is
irrelevant when $k=1$, the upshot is that our construction in $D=5$ is
equivalent to the one where the NUT charge $\ell$ is set to zero, thus
reducing to a case already considered in \cite{hawhuntay}.  In all
other dimensions $D\ge 4$, the NUT charge is a non-trivial additional 
parameter.

\section{Separability}

   The covariant Hamiltonian function on the cotangent bundle of the metrics 
(\ref{plebmetric}) is given by
\bea
{\cal H}(P_M,x^M) &\equiv& \ft12 \hat g^{MN}\, P_M\, P_N\nn\\
&=&  \fft1{2(p^2+q^2)}\, \Big[ \fft1{X}\, (P_\sigma + p^2\,
  P_\tau)^2 - \fft1{Y}\, (P_\sigma - q^2\, P_\tau)^2 + X\, P_p^2 + Y\,
  P_q^2 \Big]\nn\\
&& + \fft{\gamma}{2p^2\, q^2}\, g^{ij}\, P_i\, P_j\,.
\eea
The coordinates $\tau$ and $\sigma$ are ignorable, and their conjugate
momenta $P_\tau$ and $P_\sigma$ are constants.  The Hamiltonian-Jacobi
equation
\be
{\cal H}(\del_M S, x^M) = -\ft12 \mu^2
\ee
has separable solutions of the form
\be
S = P_\tau \, \tau + P_\sigma\, \sigma + F(p) + G(q) + W(x^i)\,,
\ee
where 
\bea
2\kappa &=&  \fft1{X}\, (P_\sigma + p^2\, P_\tau)^2 + X\, 
\Big(\fft{dF}{dp}\Big)^2 + \fft{2\gamma\, c}{p^2} + \mu^2\, p^2\,,\nn\\
-2\kappa &=& - \fft1{Y}\, (P_\sigma -q^2\, P_\tau)^2 + Y\, 
\Big(\fft{dG}{dq}\Big)^2 + \fft{2\gamma\, c}{p^2} + \mu^2\, q^2\,,\nn\\
c &=& \ft12 g^{ij}\, \fft{\del W}{\del x^i}\, \fft{\del W}{\del x^j}\,.
\label{consts}
\eea
The three quantities $\mu^2$, $c$ and $\kappa$ are separation constants, 
associated to the three mutually Poisson-commuting functions ${\cal
  H}$ and  
\bea
{\cal C} &=& \ft12 g^{ij}\, P_i\, P_j\,,\nn\\
{\cal K} &=& \fft{1}{2(p^2+q^2)}\, \Big[ \fft{q^2}{X}\, (P_\sigma+
  p^2\, P_\tau)^2 + \fft{p^2}{Y}\, (P_\sigma -q^2\, P_\tau)^2 +
q^2\, X\, P_p^2 - p^2\, Y\, P_q^2\Big] \nn\\
&&   + \ft12\gamma( \fft1{p^2} -
\fft1{q^2})\, g^{ij}\, P_i\, P_j\,. \label{calk}
\eea

    If equations (\ref{consts}) hold, then ${\cal C}$ takes the value
$c$ and ${\cal K}$ takes the value $\kappa$.  The function $W$
satisfies the Hamilton-Jacobi equation governing geodesic motion on
the $k$-dimensional Einstein manifold with metric $d\Omega_k^2$.  For
a general Einstein manifold, this is as far as one can go with finding
the geodesics.  However, in special cases, such as the sphere $S^k$,
there will be further constants of the motion.  If there are $k-1$
such constants arising from $k-1$ mutually-commuting independent
functions on the Einstein manifold, then the geodesic flow on the
$(k+4)$-dimensional Einstein manifold will be completely integrable.
Another way to say this is that while the product of two manifolds
with completely-integrable geodesics gives a new manifold with
completely-integrable geodesics, this property will not in general be
true for warped products such as we are considering here.  However, for the
very special choice of warp function $p^2\, q^2/\gamma$ which arises in
the metrics (\ref{plebmetric}), together with the simple $k$-dependent
modificications of the $X$ and $Y$ functions in (\ref{xy}), the property  
of complete integrability is maintained.     

   These complete integrability properties may be viewed as the classical 
limit of the quantum-mechanical statement that the Schr\"odinger 
equation $\hat\nabla^2\, \psi = \mu^2\, \psi$ is also 
separable.  This can be seen from the Laplacian 
\bea
\hat\nabla^2 &=& \fft1{p^2+q^2}\, \Big[ p^{-k}\, \fft{\del}{\del p}\, \left(
  p^k\, X\, \fft{\del}{\del p}\right) +
q^{-k}\, \fft{\del}{\del q}\, \left( q^k\, Y\, \fft{\del}{\del q}\right)
   \nn\\
&&\qquad \qquad + \fft1{X}\, \left(\fft{\del}{\del\sigma} + p^2\, 
\fft{\del}{\del\tau}\right)^2
 -\fft1{Y}\, \left(\fft{\del}{\del\sigma} - q^2\, 
\fft{\del}{\del\tau}\right)^2
\Big] + \fft{\gamma}{p^2 \,q^2}\, \nabla^2\,.
\eea
Multiplying $\hat\nabla^2\, \psi = \mu^2\psi$ by $(p^2+q^2)$ immediately
reveals the separability.

   Associated with ${\cal K}$ is a rank-2 symmetric St\"ackel-Killing
tensor $K^{MN}$, given by ${\cal K} = \ft12 K^{MN}\, P_M\, P_N$, where
${\cal K}$ is given in (\ref{calk}).  It satisfies the Killing-tensor
equation
\be
\hat\nabla_{(M}\, K_{NP)}=0\,,\label{skilling}
\ee
by virtue of the fact that ${\cal K}$ Poisson-commutes with the Hamiltonian.
We may also then define the second-order differential operator
\be
\hat {\cal K} = -\ft12 \hat\nabla_M( K^{MN}\, \hat\nabla_N)\,,
\ee
analogous to the operator $\hat {\cal H} = -\ft12 \hat\nabla_M(\hat
g^{MN}\, \hat\nabla_N) = -\ft12\hat\nabla^2$.  General theory
\cite{carter1} shows that $\hat{\cal K}$ and $\hat {\cal H}$ commute,
and, moreover, they obviously commute with the operator $\hat {\cal
C}\equiv -\ft12 \nabla_i(g^{ij}\, \nabla_j)=-\ft12 \nabla^2$.

   In the Carter class of four-dimensional Kerr-Taub-NUT-de Sitter
metrics, it is known \cite{carter2} that the St\"ackel-Killing tensor
$K^{MN}$ can be written as the square of a Yano-Killing 2-form
$Y_{MN}$:
\be
K_{MN}= Y_{MP}\, Y_N{}^P\,,
\ee
where $Y_{MN}$ satisfies 
\be
\hat\nabla_{(M}\, Y_{N) P}=0\,.\label{yanokilling}
\ee
This is equivalent to the statement that $\hat\nabla_M\, Y_{NP} 
= \del_{[M}\, Y_{NP]}$, or $3\hat\nabla\, Y = dY$.  It follows
straightforwardly from (\ref{yanokilling}) that $K_{MN}$ satisfies the
St\"ackel-Killing equation (\ref{skilling}).  

    In four dimensions, \ie $k=0$, one has from (\ref{calk}) that
\be
Y = q\, dp\wedge (d\tau + q^2\, d\sigma) + p\, dq\wedge 
(d\tau -p^2\, d\sigma)\,,\label{ydef}
\ee
whence
\be
{*Y} = q\, dq\wedge (d\tau-p^2\, d\sigma) - p \, dp\wedge
   (d\tau + q^2\, d\sigma)\,.
\ee
Thus one can write \cite{carter2} ${*Y}=dA$, where
\be
A= \fft{q^4}{2(p^2+q^2)}\, (d\tau - p^2\, d\sigma) -
\fft{p^4}{2(p^2+q^2)}\, (d\tau + q^2\, d\sigma)\,.
\ee

   One might wonder whether in the higher-dimensional metrics
(\ref{plebmetric}), the St\"ackel-Killing tensor we have found might also
be expressible as the square of a Yano-Killing tensor.  For a general 
dimension $D=4+k$ this looks unlikely, because there is no obvious 
2-form available on the additional $k$-dimensional Einstein manifold.
If, however, the higher-dimensional manifold is K\"ahler-Einstein, the
K\"ahler form $J_{ij}$ becomes available, and an obvious generalisation of
(\ref{ydef}) is
\be
\hat Y = Y \pm \sqrt{q^2 -p^2}\, J\,.
\ee
This is, by construction, one possible square root of the Killing tensor
$K_{MN}$.  However, a simple calculation shows that, for example, 
$\hat\nabla_\mu\, \hat Y_{jk} + \hat\nabla_j\, \hat Y_{\mu k}\ne0$,
and thus $\hat Y_{MN}$ is not a Yano-Killing tensor.  There is no
other obvious candidate for $\hat Y_{MN}$ that might yield a
generalisation of the four-dimensional Yano-Killing tensor.

   It is interesting to note that the more general class of
accelerating type-D metrics of Plebanski and Demianski \cite{plebdem}
does not admit a separation of variables, for either the
Hamiltonian-Jacobi equation or the wave equation.  These metrics are
of the form
\be
ds^2 = \fft1{(1-p\, q)^2}\, \Big[
 \fft{p^2 + q^2}{X}\, dp^2 +
\fft{p^2 + q^2}{Y}\, dq^2 +
\fft{X}{p^2 + q^2}\, (d\tau + q^2\,d\sigma)^2 -
\fft{Y}{p^2 + q^2}\, (d\tau -p^2\, d\sigma)^2\Big]\,,
\ee
where $X$ and $Y$ are certain polynomial functions of $p$ and $q$
respectively \cite{plebdem}.  The conformal prefactor $(1-p\, q)^{-2}$ 
spoils the separability of the Hamilton-Jacobi equation, except in the
massless case.  Thus the Plebanski-Demianski metrics admit conformal
St\"ackel-Killing tensors but not St\"ackel-Killing tensors in the
strict sense.

\section{Double Kerr-Schild Metric}

    It is of interest to note that the higher-dimensional metrics
(\ref{plebmetric}) may be cast in a double Kerr-Schild form.  This
is most conveniently done by analytically continuing to a real form
of the metric with signature $(2,2+k)$.  This continued metric can 
then be written in the form
\be
d\hat s^2 = d\bar s^2 + U\, (k_M\, dx^M)^2 + V \, (l_M\, dx^M)^2\,,
\label{dks}
\ee
where the fiducial ``base'' metric $d\bar s^2$ is the de Sitter
metric, and $k^M$ and $l^M$ are two linearly-independent
mutually-orthogonal affinely-parameterised null geodesic congruences:
\bea
&&  k_M\, k^M = l^M\, l_M = k^M\, l_M=0\,,\nn\\
&& k^M \, \bar\nabla_M\, k_N= l^M\, \bar\nabla_M\, l_N=0\,.
\label{kleq}
\eea
Note that the indices on $k_M$ and $l_M$ can be raised with
either $\hat g^{MN}$ or $\bar g^{MN}$.

   Specifically, the analytically-continued metric is obtained from
(\ref{plebmetric}) by sending
\bea
&&p\longrightarrow \im\, p\,,\qquad \ell\longrightarrow \im^{k-1}\,
\ell\,,\qquad \gamma\longrightarrow -\gamma\,,\nn\\
&&X \longrightarrow -\Delta_p\,,\qquad Y\longrightarrow \Delta_q\,,
\eea
resulting in the metric
\be
ds^2=\fft{q^2-p^2}{\Delta_p}\, dp^2 + \fft{q^2-p^2}{\Delta_q}\, dq^2 -
\fft{\Delta_p}{q^2-p^2}\, (d\tau + q^2\,d\sigma)^2 -
\fft{\Delta_q}{q^2-p^2}\, (d\tau+ p^2\, d\sigma)^2 + \fft{p^2\,q^2}{\gamma}
d\Omega_k^2\,,\label{pluckermet}
\ee
where
\be
\Delta_p= \gamma - \epsilon\, p^2 + \lambda\, p^4 -2\ell\, p^{1-k}\,,
\qquad
\Delta_q=-\gamma + \epsilon\, q^2 - \lambda\, q^4 - 2m\, q^{1-k}\,.
\ee

    If we now define new coordinates $\td\tau$ and $\td\sigma$ by
\be
d\td\tau = d\tau + \fft{p^2\, dp}{\Delta_p} - \fft{q^2\,
  dq}{\Delta_q}\,,\qquad
d\td\sigma = d\sigma - \fft{dp}{\Delta_p} + \fft{dq}{\Delta_q}\,,
\ee
then a straightforward calculation shows that (\ref{pluckermet}) can 
be written as (\ref{dks}), where
\bea
d\bar s^2 &=& -\fft{1}{q^2-p^2}\, \Big[\bar\Delta_p\, (d\td\tau + q^2\,
d\td\sigma)^2 - \bar\Delta_q\, (d\td\tau + p^2\, d\td\sigma)^2 \Big] 
\nn\\
&&-2(d\td\tau + q^2\,d\td\sigma)\, dp - 2(d\td\tau + p^2\, d\td\sigma)\, 
dq + \fft{p^2\, q^2}{\gamma}\, d\Omega_k^2\,,\nn\\
\bar\Delta_p &=& \gamma -\ep\, p^2 + \lambda\, p^4\,,\qquad
\bar\Delta_q = -\gamma + \ep\, q^2 -\lambda\, q^4\,,\label{ks2}\\
k_M\, dx^M &=& d\td\tau + q^2\, d\td\sigma\,,\qquad
l_M\, dx^M = d\td\tau + p^2\, d\td\sigma\,,\nn\\
U &=& \fft{2\ell\, p^{1-k}}{q^2-p^2}\,,\qquad
V = \fft{2m\, q^{1-k}}{q^2-p^2}\,.\nn
\eea
It is easily verified that $k_M$ and $l_M$ satisfy
(\ref{kleq}).  Note that as vectors, one has
\be
k^M\, \del_M = -\fft{\del}{\del q}\,,\qquad 
l^M\, \del_M = -\fft{\del}{\del p}\,.
\ee
The metric $d\bar s^2$ appearing in (\ref{ks2}) is the
$(4+k)$-dimensional de Sitter metric (since it is the Kerr-Taub-NUT de
Sitter metric with the mass and NUT parameters $m$ and $\ell$ set to
zero).  Its inverse is given simply by
\bea
\Big(\fft{\del}{\del\bar s}\Big)^2 &=& \fft1{q^2-p^2}\, \Big[ 
2(\fft{\del}{\del\td\sigma} - q^2\, \fft{\del}{\del\td\tau})\, 
\fft{\del}{\del q} - 2(\fft{\del}{\del\td\sigma} - p^2\, 
\fft{\del}{\del\td\tau})\, \fft{\del}{\del p} + 
\bar\Delta_p\,  \Big(\fft{\del}{\del p}\Big)^2 + 
\bar\Delta_q\, \Big(\fft{\del}{\del q}\Big)^2\Big]\nn\\
&& +
\fft{\gamma}{p^2\, q^2}\, \Big(\fft{\del}{\del s}\Big)^2\,,
\eea
where $(\del/\del s)^2$ is the inverse of $d\Omega_k^2$.  Note that if
one defines $h_{MN} = U\, k_M\, k_N + V\, l_M\, l_N$, so that $\hat
g_{MN} = \bar g_{MN} + h_{MN}$, then the inverse full metric is given
by $\hat g^{MN}= \bar g^{MN} - h^{MN}$, since $h_{MN}\, h^{NP}=0$.

   In a metric of single Kerr-Schild form, where $d\hat s^2 = d\bar s^2 
 + U\, (k_M\, dx^M)^2$ and $k^M$ is a null geodesic congruence, a
straightforward calculation shows that the Ricci tensor of the full metric,
written with mixed indices $\hat R^M{}_N$, is given exactly by
\cite{dergur,stkrmahohe}
\be
\hat R^M{}_N = \bar R^M{}_N - h^M{}_P\, \bar R^P{}_N + 
\ft12\bar\nabla_P \bar\nabla_N\, h^{MP} + \ft12 \bar\nabla^P\bar\nabla^M\, 
h_{NP} - \ft12 \bar\nabla^P \bar\nabla_P\, h^M{}_N\,.\label{riclin}
\ee
In other words, the ``linearised approximation'' is exact in this case.  

   In the case of the double Kerr-Schild metrics (\ref{dks}), \ie $\hat
g_{MN}= \bar g_{MN} + h_{MN}$ with $h_{MN} = U\, k_M\, k_N + V\, l_M\,
l_N$ and $k_M$ and $l_M$ satisfying (\ref{kleq}), the expression for
$\hat R^M{}_N$ in terms of $\bar R^M{}_N$ is still, of course, purely
of finite polynomial order in $h_{MN}$, because of the exact property
that $\hat g^{MN}= \bar g^{MN} - h^{MN}$.  However, the weaker
conditions satisfied by $h_{MN}$ in the double Kerr-Schild case imply
that in general, terms higher than linear order in $h_{MN}$ contribute
to $\hat R^M{}_N$.  Interestingly, however, in the specific case of
the double Kerr-Schild form (\ref{ks2}) of the higher-dimensional
Kerr-Taub-NUT-de Sitter metrics (\ref{plebmetric}), with $U$ and $V$
as in (\ref{ks2}), the linear expression (\ref{riclin}) {\it is} still
exact.  In fact, more generally we find that if one takes the functions
$U$ and $V$ to be given by
\be
U= \fft{f(p)}{q^2-p^2}\,,\qquad 
V = \fft{g(q)}{q^2 -p^2}\,,
\ee
where $f(p)$ and $g(q)$ are arbitrary functions, then the Ricci tensor
$\hat R^M{}_N$ is given exactly by (\ref{riclin}).

    It is perhaps worth remarking that one can always choose to view a
double Kerr-Schild metric of the form (\ref{dks}) as a single
Kerr-Schild metric, by including one or other of the added null terms
$U\, (k_M\, dx^M)^2$ or $V\, (l_M\, dx^M)^2$ as part of the fiducial
``base'' metric $d\bar s^2$.  Thus in or present example one can view
the higher-dimensional Kerr-Taub-NUT-de Sitter metrics
(\ref{plebmetric}) as either Kerr-de Sitter with the NUT charge added
via a Kerr-Schild term, or else as massless Kerr-Taub-NUT-de Sitter
with the mass added via a Kerr-Schild term.

\section{Euclidean-signature Metrics}

   It is also of interest to consider Einstein metrics of
positive-definite signature.  We can perform such a ``Euclideanisation''
of the metric (\ref{plebmetric}) by making the the following
analytic continuation:
\bea
&&p\rightarrow \i\, p\,, \qquad\tau \rightarrow {\rm i}\, \tau\,,\qquad
\sigma\rightarrow {\rm i}\, \sigma\,,\qquad
\ell\rightarrow {\rm i}^{k-1}\, \ell\,,\nn\\
&&X\rightarrow -\Delta_p\,,\qquad
Y\rightarrow \Delta_q\,,\qquad
\gamma\rightarrow -\gamma\,.
\eea
The metric (\ref{plebmetric}) then becomes
\be
ds^2=\fft{q^2-p^2}{\Delta_p}\, dp^2 + \fft{q^2-p^2}{\Delta_q}\, dq^2 +
\fft{\Delta_p}{q^2-p^2}\, (d\tau + q^2\,d\sigma)^2 +
\fft{\Delta_q}{q^2-p^2}\, (d\tau+ p^2\, d\sigma)^2 + \fft{p^2\,q^2}{\gamma}
d\Omega_k^2\,,\label{eucmet}
\ee
where
\be
\Delta_p= \gamma - \epsilon\, p^2 + \lambda\, p^4 -2\ell\, p^{1-k}\,,
\qquad
\Delta_q=-\gamma + \epsilon\, q^2 - \lambda\, q^4 - 2m\, q^{1-k}\,.
\ee

   It was shown recently in \cite{hassakyas} that after
Euclideanisation, the higher-dimensional Kerr-de Sitter metrics 
with a single rotation parameter that were found in \cite{hawhuntay}
yield, in a special limiting case, complete Einstein
metrics that extend smoothly onto non-singular manifolds.  Since our
more general Einstein metrics encompass those in \cite{hawhuntay},
they certainly admit the same non-singular complete metrics as special 
limiting cases.  However, the additional NUT charge parameter that 
we have in our new metrics in $D\ge 6$ provides additional
possibilities for obtaining complete, compact Einstein metrics, as
we shall now show.

   The metrics (\ref{eucmet}) are of cohomogeneity two, since the
metric functions depend on both the coordinates $p$ and $q$.\footnote{
We are not concerned here with any cohomogeneity that might be
associated with the $k$-dimensional Einstein metric $d\Omega_k^2$ if
it were not taken to be a sphere or any other homogeneous metric.  For
simplicity, and without losing any essential generality, we shall
consider
$d\Omega_k^2$ to be the round metric on $S^k$ in what follows.}  In
a compact metric, the endpoints in the ranges $p_1\le p\le p_2$ and
$q_1\le q\le q_2$ of the $p$ and $q$ coordinates will be defined by
degenerations of the metric, corresponding to collapsing of the
principal orbits.  This can occur at zeros of $\Delta_p$ or
$\Delta_q$, or at $p=0$ or $q=0$.
Although the possibility of obtaining compact
non-singular metrics of cohomogeneity two cannot be immediately
excluded, it is certainly the case that non-singularity is most easily
achieved by reducing the cohomogeneity to degree one, and so we shall
make this assumption in the discussion that follows.  The reduction of
cohomogeneity can be achieved by choosing the parameters so that the
coordinate range for either $p$ or $q$ shrinks to zero; \ie $p_1=p_2$,
or $q_1=q_2$. It turns out that for regular solutions, we should arrange
to shrink the coordinate range for $q$, by choosing the parameters so that 
$\Delta_q$ has two roots that coalesce at $q=q_0$:
\be
\Delta_q(q_0) = \Delta'_q(q_0)=0\,,\label{doubleroot}
\ee
where $\Delta_q'$ denotes the derivative of $\Delta_q$ with respect to $q$.

   It is useful to re-express $(\gamma, m, \ell,\lambda)$ in terms of 
dimensionless parameters $(g, \td m, \td\ell,L)$: 
\be
\gamma = g\, q_0^2\,,\qquad \lambda=L\, q_0^{-2}\,,\qquad
m=\td m\, q_0^{k+1}\,,\qquad
\ell=\td \ell\, q_0^{k+1}\,.
\ee
The conditions (\ref{doubleroot}) for the double root can conveniently 
be used to solve for $\epsilon$ and $\td m$ in terms of $q_0$:
\be
\epsilon=\fft{(k+3)L - (k-1) g}{k+2}\,,\qquad \td m=\td m_0=
\fft{g-L}{k+1}\,.
\ee
Moving slightly away from the case of the double root, by displacing $\td
m$ away slightly from $\td m_0$, we now define
\be
\td m=\td m_0 - \fft{\delta^2}{2c\, q_0^2}\,,\quad
q=q_0+ \delta\, \cos\theta\,,\quad
\tau=\fft{q_0^2\, c}{\delta}\, \phi \,,\quad
\sigma=\fft{2c}{q_0}\, \psi - \fft{c}{\delta}\, \phi\,,\quad
p=q_0\, r
\ee
where $c^{-1}\equiv (k+3)L-(k-1)g$, and then send $\delta \rightarrow0$.
We find that in this limit the metric becomes
\be
\fft{ds^2}{q_0^2} = 
\fft{1-r^2}{Q(r)}\, dr^2 + \fft{4c^2\, Q(r)}{1-r^2} (d\psi +
\cos\theta\, d\phi)^2 + c(1-r^2) (d\theta^2 + \sin^2\theta\, d\phi^2) +
\fft{r^2}{g}\, d\Omega_k^2\,,
\ee
where
\be
Q(r)=g - \fft{(k+3)L+ (k-1)g}{k+1}\, r^2 + L\, r^4 -
2\td \ell\, r^{1-k}\,.
\ee
This metric can be recognised as the $n=1$ special case of the class
of cohomogeneity-one Einstein metrics obtained in \cite{lpp}, having a
base Einstein-K\"ahler manifold $K_{2n}$ of dimension $2n$, with
fibres that involve a complex line bundle over $K_{2n}$ and an
additional $S^k$ warped-product factor.  The conditions for regularity of such
metrics were analysed in detail in \cite{lpp}.  Applied to our $n=1$
case where $K_2=S^2$, these results show that compact non-singular
Einstein metrics can be achieved by choosing the parameters so that
$r$ ranges either from 0 to $r_0$, where $\Delta_p(r_0)=0$, or else
by choosing the parameters such that $r$ ranges between two distinct
positive roots $r_1$ and $r_2$ of $\Delta_p$.  The former requires
$\td\ell=0$, and was in fact obtained in \cite{hassakyas} as a limit
of the single-rotation Kerr-de Sitter metrics; in this case, it is
essential for regularity that the metric $d\Omega_k^2$ be the round
metric on $S^k$.  The latter requires $\td\ell\ne0$.  It was obtained
in \cite{lpp} by starting with a cohomogeneity-one metric ansatz, but
here we have shown how it arises as a limiting case of the more
general higher-dimensional Kerr-Taub-NUT-de Sitter metrics that we
have constructed in this paper. (In this case, since the coefficient
of $d\Omega_k^2$ is everywhere non-vanishing, one can choose any
regular Einstein metric for $d\Omega_k^2$.)

\section{Conclusions}
  
   In this paper, we have studied some properties of a class of
higher-dimensional generalisations of the Kerr-Taub-NUT-de Sitter
metrics.  We have shown that they share many, but not all, of the
remarkable properties of their four-dimensional progenitors.  For
example, they admit separation of variables for both the
Hamilton-Jacobi and wave equations, and we have exhibited the
associated second-rank S\"ackel-Killing tensors.  By contrast to the
four-dimensional case, however, the S\"ackel-Killing tensor appears
not to have a Yano-Killing tensor square root.  It should be
emphasised that the separability of the higher-dimensional solutions,
which leads in many cases to completely-integrable geodesics flows, is
a rather non-trivial consequence of the detailed form of the
solutions, which is mandated by the Einstein equations.  A recent study
of separability in the higher-dimensional Kerr-de Sitter metrics showed
that this is possible at least in the special case of all rotation
parameters equal, when there is an enhanced isometry group
\cite{page2}.  In our case, the equations can be separated in
situations where there is only a single rotation parameter, together
with a NUT charge.  Previous results on separability, in the absence
of the NUT charge and cosmological constant, were obtained in
\cite{palmer}. 

   Like their four-dimensional progenitors, the metrics
(\ref{plebmetric}) may be cast in double Kerr-Schild form, which
may provide a fruitful ansatz for the further study of
higher-dimensional solutions of the Einstein equations. This is
because the solutions cast in this form may be regarded as their own 
linear approximations.  This, unlike the single Kerr-Schild ansatz,  
is not a general feature in the double Kerr-Schild case, but it does
hold for the metrics we have considered.  More generally, the double
Kerr-Schild ansatz leads to quartic non-linearity in the
Einstein equations.   

   The double Kerr-Schild form requires us to consider an
analytically-continued  form of the metric with two time directions.
Analytic continuation will also produce metrics of positive-definite
(Euclidean) signature.  This we have done, and by considering special
limiting cases, obtained complete non-singular compact Einstein
manifolds that were previously obtained in \cite{lpp}.

\end{document}